\documentclass[11pt,superscriptaddress,reprint,preprintnumbers,nofootinbib,prd]{revtex4-2}
\usepackage[utf8]{inputenc}
\usepackage[a4paper, total={7in, 9in}]{geometry}
\usepackage{amsmath}
\usepackage{amssymb}
\usepackage{comment}

\usepackage{enumerate}
\usepackage{amsmath}
\usepackage{pifont}
\usepackage{tikz}
\usepackage[compat=1.1.0]{tikz-feynman}
\tikzfeynmanset{warn luatex=false}
\usepackage{feynmf}
\usepackage{slashed}
\usepackage{braket}
\usepackage{url}
\usepackage{natbib}
\usepackage{graphicx}

\usepackage[pdftex,bookmarks,linktocpage,pdfpagelabels,plainpages=false,hyperfigures,urlcolor=brown,linkcolor=blue,citecolor=blue]{hyperref} %for hypertext links
\hypersetup{colorlinks=true}

\usepackage{mathrsfs,amssymb}  
\usepackage{cancel}
\usepackage[normalem]{ulem}

\usepackage{fontawesome}
\usepackage{xcolor}
\usepackage{xspace}
\usepackage{orcidlink}

\newcommand{\orcid}[1]{\href{https://orcid.org/#1}{#1}}

\tikzfeynmanset{
  fermion1/.style={
    /tikz/postaction={
      /tikz/decoration={
        markings,
        mark=at position 0.5 with {
          \node[
            transform shape,
            xshift=-0.25mm,
            fill,
            dart tail angle=100,
            inner sep=0.65pt,
            draw=none,
            dart
          ] { };
        },
      },
      /tikz/decorate=true,
    },
  },
  fermion2/.style={
    /tikz/postaction={
      /tikz/decoration={
        markings,
        mark=at position 0.5 with {
          \arrow{>[length=6pt, width=5pt]};
        },
      },
      /tikz/decorate=true,
    },
  },
}

\makeatletter
\def\@hangfrom@section#1#2#3{\@hangfrom{#1#2}#3}%\MakeTextUppercase{#3}}%
\def\@hangfroms@section#1#2{#1#2}%\MakeTextUppercase{#2}}%
\makeatother

\begin{document}

\preprint{CETUP2025-016}

\title{Enhanced active-sterile neutrino polarizability at the intensity frontier}

\author{Julia Gehrlein}

\email{julia.gehrlein@colostate.edu}
\thanks{\orcid{0000-0002-1235-0505}}
\affiliation{Physics Department, Colorado State University, Fort Collins, CO 80523, USA}

\author{Anil Thapa}

\email{a.thapa@colostate.edu}
\thanks{\orcid{0000-0003-4471-2336}}
\affiliation{Physics Department, Colorado State University, Fort Collins, CO 80523, USA}

\author{Adrian Thompson}
\email{a.thompson@northwestern.edu}
\thanks{\orcid{0000-0002-9235-0846}}
\affiliation{Northwestern~University,~Evanston,~IL~60208,~USA}

\begin{abstract}
Electromagnetic probes of neutrinos can provide insights into physics beyond the Standard Model. Among the possible electromagnetic interactions of neutrinos is neutrino polarizability, a dimension-7 effective operator that couples two neutrinos to two photons. In this manuscript, we study a realization of the neutrino polarizability operator in which one of the active neutrinos is replaced by a sterile neutrino.  We derive new constraints on this active-sterile neutrino polarizability from its contribution to neutrino-nucleus scattering with a single photon in the final state at neutrino experiments. We show that a realization of this operator via a light mediator can explain the MiniBooNE low-energy excess while remaining consistent with other experimental constraints. Finally, we comment on additional model realizations of this higher-dimensional operator.
\end{abstract}

\maketitle

\section{Introduction}
Neutrinos could have interactions beyond the Standard Model (SM). To remain agnostic about the underlying physics, these new interactions are commonly parametrized via effective operators. Indeed, the well studied case of new neutrino interactions with other SM fermions is parametrized in the framework of neutrino nonstandard interactions  (NSI) \cite{Wolfenstein:1977ue} which couples two neutrino fields to two fermion fields, resulting in an effective four-fermion operator. By contrast, effective couplings of neutrinos to SM gauge bosons are less well studied; so far, the focus has primarily been on neutrino-photon couplings.

The two-photon neutrino vertex, also referred to as  neutrino polarizability, is one such dimension-7 effective interaction between the scalar/pseudoscalar neutrino current and two photons. Originally studied by Gell-Mann~\cite{Gell-Mann:1961pqd} and Levine~\cite{Levine1967}, it was shown to vanish except in the case of massive neutrinos. Calculations of this vertex within the electroweak theory were performed~\cite{Karl:2004bt,LiuPRD, Crewther:1981wh}, with many applications to astrophysical environments in mind. For example, the $\nu\nu\gamma\gamma$ vertex could contribute as a stellar cooling mechanism, primarily through $\gamma \gamma \to \nu \bar{\nu}$ annihilation in stellar cores~\cite{Levine1967,dodelson-feinberg-prd}, but also through induced photon decays $\gamma \to \gamma \nu \bar{\nu}$ and fusion $\gamma \gamma \to \nu \bar{\nu}$ in strong astrophysical magnetic fields~\cite{Dicus:1993iy,Dicus:1997rw,Dicus:2000cz,Chistyakov:2002re}. Laboratory probes using laser polarizability~\cite{Mohammadi:2013ksa} and birefringence~\cite{Petropavlova:2022spq} have also been considered, in addition to related operators with one extra photon attached in the context of astrophysical environments~\cite{Ahmadiniaz:2023udn}. In most of these contexts, the polarizability is highly suppressed by the small size of the neutrino masses, and hard to observe. However, recent efforts have realized specific models that enhance the polarizability~\cite{Bansal:2022zpi,Gehrlein:2025tko,Carey:2025oxt}.

Generating such effective interactions in a UV complete theory is a nontrivial task, as it requires respecting $SU(2)_L$ gauge invariance. Operators involving active, left-handed neutrinos typically induce correlated interactions of the charged leptons in the same $SU(2)_L$ doublets, which are tightly constrained (see Ref.~ \cite{Coloma:2024ict} for the case of NSI). These constraints can be alleviated if one of the active neutrinos is replaced by a sterile neutrino below the electroweak scale \cite{Li:2020wxi,Li:2021tsq}.
A common way to parametrize effective interactions among SM fields below the weak scale is the LEFT (low-energy effective field theory) framework, which extends to the $\nu$LEFT framework in the presence of a light sterile neutrino. In this setup, a dimension-7 $\nu$LEFT operator describing active-sterile neutrino polarizability can be written as
\begin{equation}
    \mathscr{L}_{\rm pol} =  \dfrac{C_{ij}}{\Lambda^3}\overline{N}_{j} \nu_{i} F^{\mu\nu} \widetilde{F}_{\mu\nu} \, ,
    \label{eq:effopt}
\end{equation}
where $\widetilde{F}_{\mu \nu} = \frac{1}{2} \epsilon_{\mu \nu \rho \sigma} F^{\rho \sigma}$ is the dual electromagnetic field strength tensor, $C_{ij}$ is a Wilson coefficient, and $\Lambda$ is the scale of new physics. This Wilson coefficient will govern the scattering with leptons and protons, atoms, and quarks across  the different kinematic regimes, see sec.~\ref{sec:pheno} for more details on the implementation.
A similar dimension-7 operator also exists with $F^{\mu\nu}F_{\mu\nu}$. Before electroweak symmetry is broken the operator of Eq.~\eqref{eq:effopt} arises from a dimension-8 operator of the form $(\bar N_{i} H L_j)\, B_{\mu\nu}\tilde B^{\mu\nu}$. Analogous operators with $W_{\mu\nu}$ also exist. Here $N$ is the sterile state with the Majorana mass given by
\begin{equation}
    {\mathscr{L}}_{\rm mass} = \frac{1}{2} m_N  \bar{N}^c N \, .
    \label{eq:sterilemass}
\end{equation} 

\begin{figure}[t!]
    \centering
    \begin{tikzpicture}
              \begin{feynman}
         \vertex (c);
         \vertex [left = 1.2cm of c] (i1) {\(N\)};
         \vertex [right = 1.2cm of c] (i2) {\(\nu\)};
         \vertex [above right = 1.2cm of c] (g1) {\(\gamma\)};
         \vertex [above left = 1.2cm of c] (g2) {\(\gamma\)};

         \node[draw, fill, rectangle, inner sep=2pt] at (c) {};
         
         \diagram* {
           (c) -- [boson] (g1),
           (c) -- [boson] (g2),
           (i1) -- [fermion1] (c) -- [fermion1] (i2)
         };
        \end{feynman}
       \end{tikzpicture}
       
    \caption{Polarizability effective vertex between two photons, one active neutrino, and one sterile neutrino.}
    \label{fig:pol_feynman}
\end{figure}
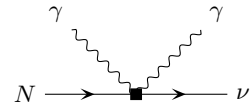

The operator in Eq.~\ref{eq:effopt} with a massive sterile state is illustrated in Fig.~\ref{fig:pol_feynman}.
This interaction leads to new contributions to neutral-current neutrino-nucleus scattering with a mono-photon in the final state (NC$1\gamma$), as well as to a new sterile neutrino decay channel into an active neutrino and two photons which does not depend on the active-sterile mixing angle. Due to the similarity of the first process to the commonly studied inverse Primakoff scattering process, we refer to the $\nu \mathcal{A}\to \mathcal{A}\,N \gamma $ process as neutrino-induced inverse Primakoff  ($\nu$IIP) scattering.
So far, this active-sterile polarizability operator has not been studied, whereas constraints on the operator with two active neutrinos exist \cite{Bansal:2022zpi,Carey:2025oxt,Gehrlein:2025tko}. Compared to the case with only active neutrinos, the active-sterile neutrino polarizability operator leads to different final state kinematics in neutrino-nucleus scatterings. Crucially, this process exhibits a final state with a minimally recoiling nucleus (provided the incoming active neutrino energy is less than around a GeV) and a single photon with a broad energy spectrum. At neutrino beam experiments, this process differentiates from standard neutrino background events like NC$1\gamma 0 p$ via $\Delta$ resonance production and other channels. At MeV neutrino experiments like reactors, the final state photon inherits most of the incoming neutrino energy, supporting a 1-10 MeV electromagnetic signal that can be searched for at detectors with nominal threshold detection energies and competes only with radioisotope backgrounds.

An ideal place to study neutrino polarizability is at experiments with an intense neutrino beam and a sensitive photon detector which can distinguish an electromagnetic cascade originating from electron interactions from the signature of a single photon. Accelerator neutrino experiments with LAr detectors such as the suite of detectors at the SBN neutrino program at Fermilab (MicroBooNE, SBND, and ICARUS) \cite{Machado:2019oxb}, and the future DUNE detectors \cite{DUNE:2020ypp,DUNE:2021tad},   reactor experiments like JUNO-TAO \cite{JUNO:2020ijm} and PROSPECT-II \cite{PROSPECT:2021jey}, or experiments using neutrinos from pion decay at rest, e.g. CCM~\cite{CCM:2021jmk,CCM:2021leg,CCM:2023itc}, JSNS$^2$ and JSNS$^2$-II~\cite{JSNS2:2017gzk,JSNS2:2020hmg}, and COHERENT (COH-Ar-750)~\cite{COHERENT:2022nrm}, will realize these requirements.  

There are also existing searches for the NC$1\gamma$ final state from NOMAD \cite{NOMAD:2011gyy},  T2K \cite{T2K:2019odo}, MicroBooNE \cite{MicroBooNE:2025ntu}, and MiniBooNE \cite{MiniBooNE:2018esg,MiniBooNE:2020pnu}. In particular, MiniBooNE cannot distinguish electrons from photons. Interestingly, the MiniBooNE experiment has reported a significant excess (4.8$\sigma$) of electromagnetic-like events at low energy \cite{MiniBooNE:2018esg,MiniBooNE:2020pnu}, which cannot, to the best of our knowledge, be explained with SM physics alone \cite{Brdar:2021ysi,Kelly:2022uaa}, leaving the possibility of new physics explanations open. Additionally, MicroBooNE has recently observed an excess of photon-like events at $2.2\sigma$~\cite{MicroBooNE:2025ntu}, providing mild supporting evidence for scattered mono-photon explanations of the anomaly~\cite{Harvey:2007ca,Harvey:2007rd,Alvarez-Ruso:2021dna,Abdallah:2022grs,CCM:2023itc,Ioannisian:2025bro,Dutta:2025fgz,Kamp:2023hai,Abdullahi:2023ejc}.  
In this manuscript, we put forward a new solution to the MiniBooNE anomaly which is in agreement with all other constraints. 
Furthermore, we contrast the active-sterile polarizability operator with existing mono-photon data and projected sensitivities from these experiments to derive the corresponding bounds. 

The outline of this manuscript is as follows: 
We first discuss the phenomenology of  the $\nu$IIP  process in sec.~\ref{sec:pheno} before we  present our analysis methodology in sec.~\ref{sec:analysis}.  In sec.~\ref{sec:eff_operator} we derive constraints on the  effective operator.
Then we study a possible UV realization of the neutrino polarizability operator with a sterile neutrino in sec.~\ref{sec:MB} and show that this scenario can explain the MiniBooNE low-energy excess. We subsequently  discuss other UV completions in sec.~\ref{sec:models}, before we conclude in sec.~\ref{sec:conclusions}.
 
\section{Phenomenology}
\label{sec:pheno}
We begin by discussing the phenomenology of the $\nu$IIP process $\nu\mathcal{A} \to  \mathcal{A}\,N\gamma$ and the impact of the sterile neutrino masses on the observables. At low momentum transfer of MeV's and below, the scattering is coherent with the entire atom $\mathcal{A}$ and enhanced by the atomic number $Z^2$ via the atomic form factor, for which we use the combined dipole form~\cite{RevModPhys.46.815} for the screened atomic potential with the Helm parametrization of the nuclear charge density, see e.g. Ref.~\cite{Engel:1992bf}~\footnote{For a discussion, see Appendix B of Ref.~\cite{Brdar:2025hqi}.}. The main experimental signature of the coherent $\nu$IIP is therefore a single photon in the final state with minimal atomic recoil and the outgoing neutrino escaping undetected. The key observables are hence the photon energy and its angular distribution with respect to the incoming neutrino beam. The sterile neutrino mass $m_N$ primarily controls how the incoming neutrino energy is shared between the photon and the final state neutrino. In the limit $m_N=0$ (corresponding to a purely active neutrino in the final state), the photon can carry away an ${\cal O}(1)$ fraction of the incoming neutrino energy, and the spectrum $E_\gamma/E_\nu$ is comparatively hard. Once $m_N \neq 0$, part of the incoming energy is shifted to lower $E_\gamma$, and the available phase space shrinks.

\begin{figure}
    \centering
    \includegraphics[width=0.48\textwidth]{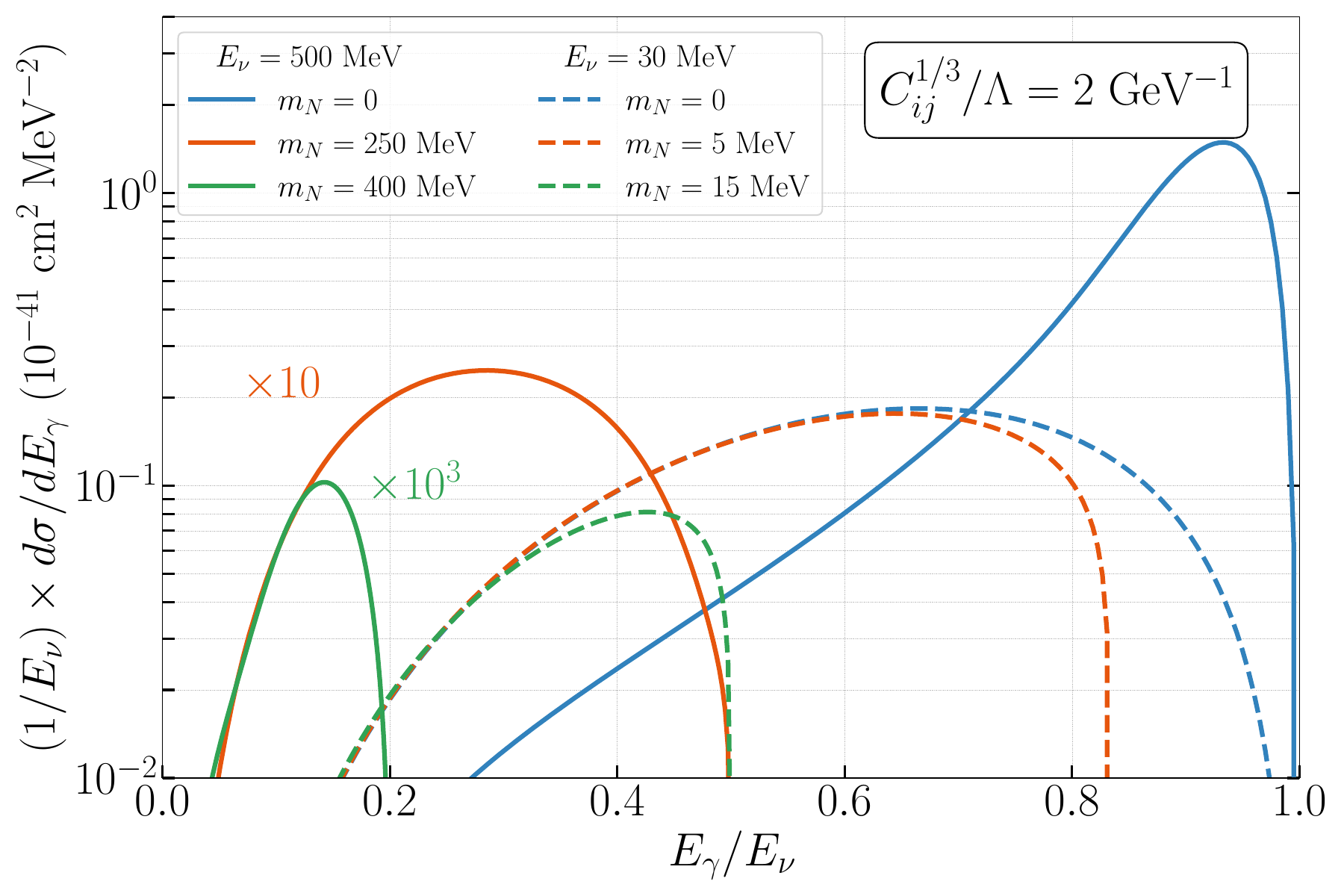}
    \caption{Differential cross section, divided by the incoming neutrino energy $E_\nu$, for the coherent $\nu_a \mathcal{A} \to N \mathcal{A} \gamma$ scattering off an atomic target ${\cal A}$ with respect to the final state photon energy $E_\gamma$, for massless incoming neutrinos and benchmark choices for the heavy neutrino mass $m_{N}$. For low energies, the differential cross section decreases and shifts to lower photon energies as $m_{N}$ approaches threshold due to kinematic phase space alone (dashed lines). At higher energies, as $m_{N}$ grows the increased momentum transfer suppresses the coherence via the nuclear form factor; hence, the $m_{N}=250, \, 400$ MeV cases are upscaled by $10$ and $10^3$, respectively.}
    \label{fig:atomic_diffxs}
\end{figure}

\begin{figure*}
    \centering
    \includegraphics[width=1.0\textwidth]{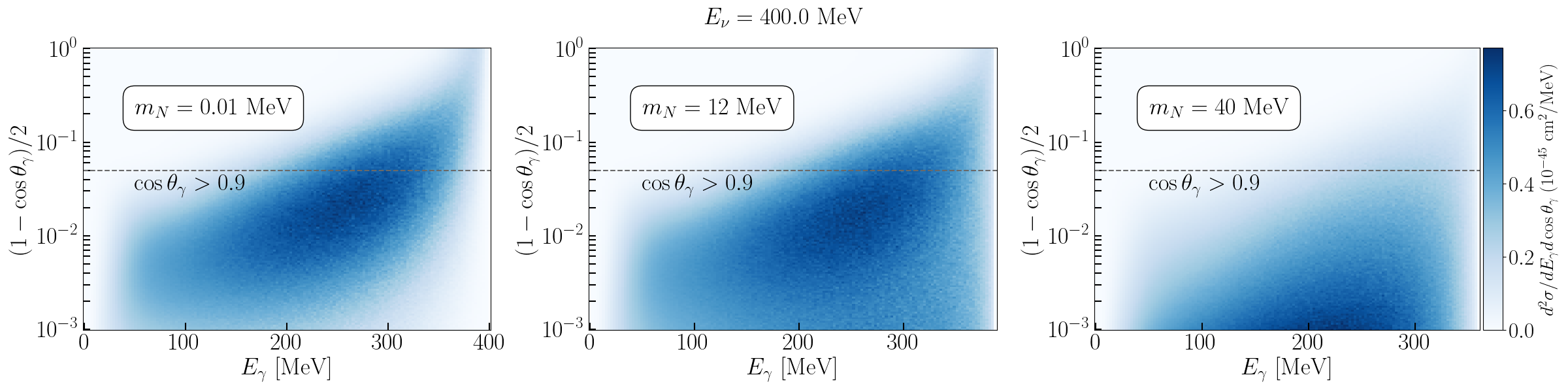}
    \caption{Heat maps of the two-dimensional differential $\nu$IIP coherent cross section, in the heavy EFT picture (Eq.~\ref{eq:effopt}), over the outgoing photon energy $E_\gamma$ and cosine of the angle with respect to the parent neutrino direction $\cos\theta_\gamma$; the vertical axis is shown over $(1-\cos\theta_\gamma)/2$ to instead show the very-forward region. Below the gray dashed line indicates where $\cos\theta_\gamma > 0.9$ corresponding to the very-forward region.}
    \label{fig:atomic_diffangle}
\end{figure*}

This behavior is illustrated in Fig.~\ref{fig:atomic_diffxs}, which shows the differential cross section, normalized by the incoming neutrino energy, as a function of $E_\gamma/E_\nu$ for the different benchmark choices of $m_N$ and the neutrino energy at $E_\nu=30$ MeV and $E_\nu=500$ MeV. For $E_\nu = 30$ MeV (dashed curves), a nonzero $m_N$ shifts both the endpoint and the peak of the spectrum to lower $E_\gamma/E_\nu$ and reduces the rate through phase-space suppression. At $E_\nu = 500$ MeV (solid curves), increasing $m_N$ further suppresses the cross section by pushing the process to larger momentum transfer and reducing coherence. Hence the $m_N = 250$ MeV and $m_N=400$ MeV curves are rescaled by factors of $10$ and $10^3$, respectively, for visibility.

The corresponding angular information is shown in Fig.~\ref{fig:atomic_diffangle}, where the two-dimensional differential $\nu$IIP coherent cross section is displayed as a function of the photon energy $E_\gamma$ and the angle $\theta_\gamma$ with respect to the parent neutrino direction. The vertical axis is plotted in terms of $(1-\cos\theta_\gamma)/2$ in order to emphasize the very-forward region. As expected for a coherent process mediated by a $t$-channel photon, the emission is strongly forward-peaked: most of the rate lies below the gray dashed line corresponding to $\cos\theta_\gamma > 0.9$. The highest-energy photons are the most tightly collimated along the beam. Taken together, these features imply that, relative to the $m_N \to 0$ case, a nonzero sterile-neutrino mass leads to a softer photon spectrum with the same characteristic forward peaking, a pattern that will be crucial for assessing the experimental sensitivity in the setups considered below.

\begin{figure}[th]
    \centering
    \begin{tikzpicture}
    \begin{feynman}
        \vertex (v1);
        \vertex [left=1.5cm of v1](i1) {\(\nu\)};
        \vertex [above right=1.2cm of v1] (f1) {\(N\)};
        \vertex [right=1.2cm of v1] (f2) {\(\gamma\)};
        \node [below=1.0cm of v1, crossed dot, label={[left=5pt]{\(\mathcal{A}\)}}] (b);

        \node at (3.0, -0.5) {Coherent};
        
        \diagram* {
        (i1) -- [fermion1] (v1) -- [fermion1] (f1),
        (v1) -- [boson] (f2),
        (v1) -- [boson] (b),
        };
    \end{feynman}
    \end{tikzpicture}
    \begin{tikzpicture}
    \begin{feynman}
        \vertex (v1);
        \vertex [left=1.5cm of v1](i1) {\(\nu\)};
        \vertex [above right=1.2cm of v1] (f1) {\(N\)};
        \vertex [right=1.4cm of v1] (f2) {\(\gamma\)};
        \vertex [below=1.0cm of v1] (b);
       
        \vertex [right=1.5cm of b] (p2) {\(f\)};
        \vertex [left=1.5cm of b] (p1) {\(f\)};

        \node at (3.0, -0.5) {Incoherent};
        
        \diagram* {
        (i1) -- [fermion1] (v1) -- [fermion1] (f1),
        (v1) -- [boson] (f2),
        (v1) -- [boson] (b),
        (p1) -- [fermion1] (b) -- [fermion1] (p2),
        };
    \end{feynman}
    \end{tikzpicture}
    \begin{tikzpicture}
    \begin{feynman}
        \vertex (v1) at (-0.2,0);
        \vertex [left=1.5cm of v1](i1) {\(\nu\)};
        \vertex [above right=1.2cm of v1] (f1) {\(N\)};
        \vertex [right=1.2cm of v1] (f2) {\(\gamma\)};
        \vertex [blob] (b) at (-0.2,-1) {};
        \vertex [right=1.5cm of b] (p2);
        \vertex [left=1.7cm of b] (p1) {\(u,d\)};

        \vertex (o1) at (1.6, -1.2);
        \vertex (o2) at (1.6, -0.8);
        \vertex (o3) at (1.6, -1.0);

        \node at (3.0, -0.5) {DIS};
        
        \diagram* {
        (i1) -- [fermion1] (v1) -- [fermion1] (f1),
        (v1) -- [boson] (f2),
        (v1) -- [boson] (b),
        (p1) -- [fermion1] (b),
        (b) -- (o1),
        (b) -- (o2),
        (b) -- (o3),
        };
    \end{feynman}
    \end{tikzpicture}
    \caption{Neutrino-induced inverse Primakoff ($\nu$IIP) coherent scattering on atomic targets (top), incoherent scattering on individual fermions $f=e^-, p$ (middle), and deep inelastic scattering (bottom) each become dominant in the soft ($q^2 \lesssim 0.1$ GeV$^2$), high energy ($q^2 \sim$ GeV$^2$), and very high energy regimes, respectively.}
    \label{fig:vIIP_diagrams}
\end{figure}
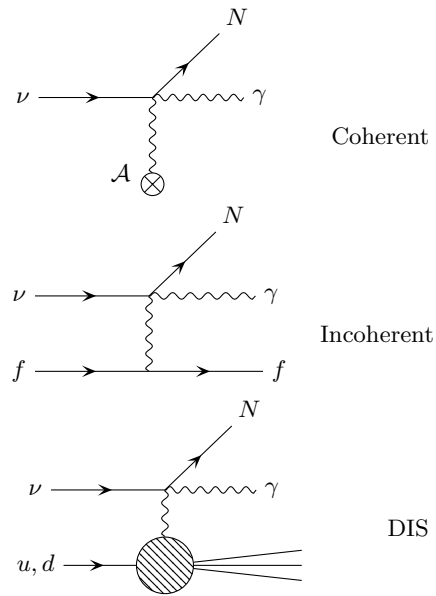

Since the incoming neutrino energies vary significantly across different experiments --  ranging from the MeV scale at reactor and pion-decay-at-rest experiments to the multi-GeV at accelerator neutrino experiments -- we
consider different kinematic regimes for the scattering based on the momentum transfer $q^2 = - (p_{\cal A}^{\rm out} - p_{\cal A}^{\rm in})^2$: coherent ($q^2 < 0.1\ {\rm GeV}^2$), incoherent ($0.1\ {\rm GeV}^2 < q^2 < 1.8\ {\rm GeV}^2$), and deep inelastic scattering (DIS) ($q^2 > 1.8\ {\rm GeV}^2$), represented in Fig.~\ref{fig:vIIP_diagrams}. To calculate the cross section in the different regimes, we implement our model in the {\tt FeynRules} package \cite{Christensen:2008py} and compute the cross-sections at the parton level for all regimes using the {\tt MadGraph5} event generator \cite{Alwall:2014hca}. In the coherent regime we use the Helm form factor \cite{Helm:1956zz,Lewin:1995rx} to parametrize the scattering, in the incoherent regime we use the dipole form factor. Finally, in the DIS regime  we use the parton distribution functions (PDFs) of the proton from the NNPDF2.3 PDF set \cite{Ball:2013hta}.  
We then sum the contributions from these regions incoherently to obtain the final cross section. 

See the  appendix of \cite{Gehrlein:2025tko} for more details on the implementation of the scattering cross section in the different regimes.
We assume  that the same Wilson coefficient defined in eq.~\eqref{eq:effopt} governs the interactions in the different kinematic regimes. To map this operator to  the different scattering regimes, we use different form factors (Helm form factor in the coherent regime, dipole form factor in the incoherent regime).

Detectors with excellent electromagnetic energy resolution are required to distinguish between different sterile–neutrino masses in the NC$1\gamma$ channel. For $E_\gamma \sim 50$--$300~\text{MeV}$, a fractional energy resolution at the level of $\sim 10$--$20\%$ (corresponding to a few MeV) and an angular resolution of a few degrees are sufficient to resolve the distortions in the photon spectrum and angular distribution induced by $m_N$. Such performance is already achieved in existing LArTPCs: a dedicated MicroBooNE study of $\pi^0\to\gamma\gamma$ showers reports a Gaussian photon–energy resolution of $8$--$12\%$ (with a $68\%$ containment width of $15$--$20\%$) over $E_\gamma \simeq 50$--$300~\text{MeV}$, and a 3D angular resolution that peaks at $2.7^\circ$, with about $60\%$ of photons reconstructed within $10^\circ$ of the true direction~\cite{MicroBooNE:2019rgx}. Similar ${\cal O}(10\%)$ energy and degree–level angular resolutions for electromagnetic showers are adopted as design targets in DUNE and SBN LArTPC studies~\cite{DUNE:2020ypp,Machado:2019oxb,Caratelli:2022llt}.

Other processes, such as neutrino upscattering into a heavy state followed by its radiative decay $N\to \nu \gamma$ via a transition magnetic moment~\cite{Gninenko:2009ks,Gninenko:2010pr,Gninenko:2012rw,Radionov:2013mca,Masip:2012ke,Schwetz:2020xra,Magill:2018jla,Fischer:2019fbw,Kamp:2022bpt,Alvarez-Ruso:2021dna,Vergani:2021tgc} can also lead to a single photon final state however with potentially different photon kinematics. In contrast, models in which a new scalar is produced and subsequently decays into two photons predict a two-photon final state~\cite{Datta:2020auq,Dutta:2021cip,Dutta:2025fgz,Chang:2021myh}. Both classes of models have been invoked as possible explanations of the MiniBooNE excess~\cite{Acero:2022wqg,Abdullahi:2023ejc}.
In the SM, the NC$1\gamma$ final state is also expected, arising from the de-excitation of a nuclear resonance in NC scattering \cite{Wang:2013wva}. However, unlike the $\nu$IIP with only active neutrinos, the active-sterile polarizability operator studied here produces a sterile neutrino in the final state and therefore does not interfere with the SM NC$1\gamma$ amplitude.

In addition to the $\nu$IIP process, the neutrino polarizability operator with a sterile neutrino also leads to a new decay channel of the heavy state, namely $N\to \nu \gamma \gamma$.
The decay rate is \cite{Larios:2002gq} 
\begin{equation}
    \Gamma(N \to \nu \gamma\gamma)
= \frac{m_{N}^{7}}{1920\,\pi^{3}}\;
\left|\frac{C_{ij}}{\Lambda^3} \right|^{2},
\end{equation}
where $C_{ij}/\Lambda^3$ is the coefficient of the dimension-7 operator in Eq.~\eqref{eq:effopt}. This strong $m_N^7$ dependence implies that, for the values of $C_{ij}/\Lambda^3 \lesssim 10^{-5}~\text{GeV}^{-3}$ relevant to our analysis (see Fig.~\ref{fig:constraint_wilson}), the decay is typically very slow with lifetime
\begin{equation}
    \tau_N \gtrsim 4\times 10^{-10}~\text{s}
    \left(\frac{1~\text{GeV}}{m_N}\right)^{7}.
\end{equation}
so that $m_N = 1~\text{GeV}$ corresponds to a decay length $c\tau_N \sim 0.1~\text{m}$, while for $m_N = 0.1~\text{GeV}$ the lifetime increases and the corresponding decay length is $c\tau_N \sim 10^{6}~\text{m}$. The experimental signature of this decay is a pair of photons arising from a common vertex, and depending on the UV completion, the total decay rate could be enhanced, potentially making multi-photon events observable.  

There are also additional constraints on this operator from solar-neutrino scattering at XENONnT and Borexino~\cite{XENON:2022ltv,Borexino:2017fbd}, as well as from
cosmological observables, stellar cooling, and meson decays. Note, however, that these bounds were derived for the case of neutrino polarizability with active neutrinos only. Since the final state neutrino is not detected, these constraints
only apply to sterile-neutrino masses below the maximum neutrino energy relevant for the corresponding experiments. For further details, see the appendix of
Ref.~\cite{Gehrlein:2025tko} and the discussion in Ref.~\cite{Bansal:2022zpi}. Lastly, the active-sterile polarizability may also support processes taking place inside accelerator targets or reactor cores via the electromagnetic showers, e.g. $\gamma \mathcal{A} \to \bar{\nu} N \mathcal{A}$. For the scope of this work, we neglect such processes, although they may prove relevant for a future work, focusing only on the phenomenology of the active-sterile polarizability in the detector.

\section{Methodology}
\label{sec:analysis}
We study the process $\nu {\cal A}\to {\cal A} \, N \gamma$ with a single photon in the final state, where ${\cal A}$ represents the atomic target (either an atomic target in the coherent regime or an individual nucleon/parton at higher momentum transfer). To obtain the most sensitivity to the effects of the neutrino polarizability operator, we   
consider this process at neutrino experiments with an intense neutrino beam and a large detector.  The  sterile neutrino mass that can be probed is kinematically limited by the maximum incoming neutrino energy in a given experiment. 

As an example of a high-energy accelerator neutrino experiment with an existing mono-photon data set, we study the NOMAD experiment \cite{NOMAD:2011gyy}. An intense flux of GeV-energy neutrinos is also produced at the Booster Neutrino Beam (BNB) at Fermilab, which can be detected at MiniBooNE and at the SBN detector ICARUS, MicroBooNE, and SBND \cite{Machado:2019oxb}, and the LBNF neutrino beam will provide an even more intense neutrino flux to be detected at the DUNE near detector \cite{DUNE:2021tad}.  MiniBooNE's Cherenkov detector cannot distinguish an electron shower from a photon shower, so its electromagnetic shower data set~\cite{MiniBooNE:2020pnu} can be directly reinterpreted in terms of NC$1\gamma$ production. The T2K near detector~\cite{T2K:2019odo} has also reported results on this final state.
At sub-GeV energies, reactor and pion-decay-at-rest experiments provide the most intense neutrino fluxes. As concrete examples, we will analyze JUNO-TAO~\cite{JUNO:2020ijm} and PROSPECT-II at HFIR~\cite{PROSPECT:2021jey}.  Further details of all these experimental setups used in our analysis are briefly summarized in the appendix. 

To obtain the predicted photon spectrum as a function of the photon energy and solid angle in a fixed proton-on-target (POT) neutrino experiment, we compute 
\begin{equation}
\frac{d^2 N}{d E_\gamma d\Omega_\gamma}= \mathscr{E} \int \epsilon\left(E_\gamma\right) \phi_{\rm tot}(E_\nu) \frac{d^2 \sigma(E_\nu)}{d E_\gamma d\Omega_\gamma} d E_\nu \, ,
\end{equation}
where the exposure factor $\mathscr{E} = N_T \times N_{\rm POT}$, with $N_T$ the number of target atoms in the fiducial detector volume and $N_{\rm POT}$ the number of protons on the beam target. The flux summed over flavor components is $\phi_{\rm tot}(E_\nu)$ (in units of cm$^{-2}$~GeV$^{-1}$~POT$^{-1}$), $d^2\sigma/(dE_\gamma d\Omega_\gamma)$ is the total double-differential cross section per target atom (cm$^{2}$~MeV$^{-1}$~sr$^{-1}$) incorporating coherent, incoherent, and DIS contributions, and $\epsilon (E_\gamma)$ is the photon detection efficiency. For reactor antineutrino experiments, we replace the exposure factor with $\mathscr{E} = \tau \times N_T$ where $\tau$ is the exposure time in seconds. The flux in that case is taken to be isotropic,
\begin{equation}
    \phi_{\rm tot}(E_\nu)
    = \frac{W}{4\pi L^2}\,\frac{dN_\nu}{dE_\nu}\,,
\end{equation}
where $L$ is the core-detector baseline, $W$ is a reactor power rescaling factor, and $dN_\nu/dE_\nu$ is the differential antineutrino energy spectrum in units of MeV$^{-1}$ s$^{-1}$.
To obtain the differential cross section $d^2 \sigma/(d E_\gamma d E_\nu)$, we implement effective operator in the {\tt FeynRules} package~\cite{Christensen:2008py} and generate the corresponding model file, which we use to compute the parton-level cross sections in all kinematic regimes with the {\tt MadGraph5\_aMC@NLO} event generator~\cite{Alwall:2014hca}. 
 
For experiments with existing NC$1\gamma$ data, we perform a chi-squared ($\chi^2$) analysis using the number of observed and expected background events, taking into account both statistical and systematic uncertainties for each experiment.  For MiniBooNE, we treat the neutrino and the antineutrino runs independently, compute the individual contributions $\chi^2_\nu$ and $\chi^2_{\bar \nu}$, and then obtain constraints on the parameter space by minimizing the sum $\chi^2_\nu + \chi^2_{\bar \nu}$. All of these experiments provide both the photon energy and angular spectrum. Since the correlations between the two observables are not publicly available, we derive independent constraints from each distribution. We find that the photon energy spectrum yields a slightly stronger bound, and we adopt it for our main results. 
To derive forecasted constraints from upcoming experiments, we 
assume that no excess events are observed and impose the condition that the total number of predicted signal events remain below 2.7 at 90\% C.L assuming Poisson statistics.

\section{Results}

\subsection{Constraints on the effective operator}
\label{sec:eff_operator}

\begin{figure}
    \centering \includegraphics[width=0.45\textwidth]{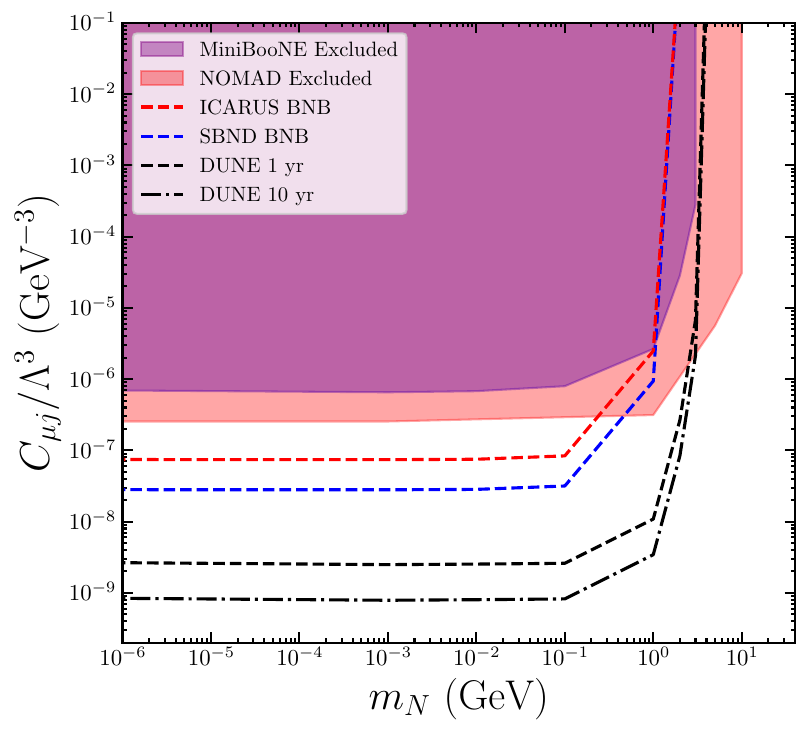} \\
    \includegraphics[width=0.45\textwidth]{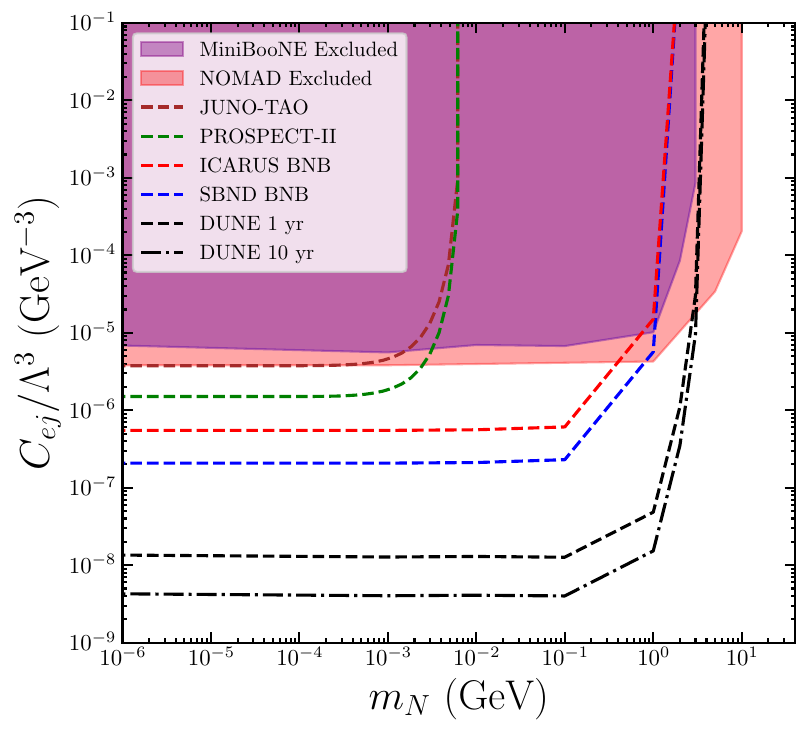}
    \caption{Exclusion region and future sensitivity from different neutrino experiments in the plane of the mass of the sterile state and the effective couplings to muon neutrinos and photons $C_{\mu j}/\Lambda^3$ (upper plot) and electron neutrinos and photons $C_{e j}/\Lambda^3$ (lower plot). Future experimental projections are shown as dashed lines. }
    \label{fig:constraint_wilson}
\end{figure}

We show our results on the active-sterile polarizability operator of Eq.~\eqref{eq:effopt} in Fig.~\ref{fig:constraint_wilson}. The upper plot shows the case in which the operator couples to muon neutrinos in the $(m_N,\,C_{\mu j}/\Lambda^3)$ plane, while the lower panel displays the corresponding results for electron neutrinos, $C_{e j}/\Lambda^3$. In each figure, shaded regions indicate the currently excluded parameter space, and the dashed lines show the projected sensitivity reach of  future experiments. For clarity, only the strongest bounds are shown in each case. 

For the muon-flavor operator (top figure), the strongest current limit comes from NOMAD, which excludes couplings larger than $C_{\mu j}/\Lambda^3 \sim 5\times 10^{-7}~\text{GeV}^{-3}$ for sterile masses
below $m_N \simeq 1~\text{GeV}$. MiniBooNE provides a complementary bound at
lower masses, and both exclusions weaken rapidly once $m_N$ approaches the
maximum neutrino energy available at each experiment, where on-shell production
is no longer possible. The projected sensitivities of the SBND and ICARUS from the BNB flux improve current existing accelerator limits by roughly an order in magnitude over most of the parameter space. The DUNE near detector, shown for one and ten years of data taking, is expected to probe down to
$C_{\mu j}/\Lambda^3 \sim 10^{-8}$–$10^{-9}~\text{GeV}^{-3}$ for $m_N \lesssim \mathcal{O}(1~\text{GeV})$, corresponding to an improvement of up to three orders of magnitude relative to the current NOMAD bound.

The electron–flavor operator (bottom figure) exhibits a similar structure,
with NOMAD again providing the most stringent existing accelerator constraint,
at the level of $C_{e j}/\Lambda^3 \sim 5\times 10^{-6}~\text{GeV}^{-3}$ for
$m_N \lesssim 1~\text{GeV}$. In this case, however, there are additional constraints from reactor experiments. Future reactor experiments like JUNO-TAO and PROSPECT-II will provide competitive  yet slightly weaker constraints than upcoming accelerator neutrino experiments. As in the muon channel, the projected sensitivities of SBND, ICARUS, and especially the DUNE near detector extend the reach in $C_{e j}/\Lambda^3$ by several orders of magnitude for $m_N$ below the respective kinematic thresholds. We find that SBND and ICARUS provide stronger sensitivity than MiniBooNE, despite MiniBooNE’s larger detector mass due to the  LArTPC capabilities leading to an increased 
    efficiency, and the larger number of nuclei in argon compared to carbon which increases the coherent cross section. We omit showing the constraints from MicroBooNE as they are weaker than the MiniBooNe constraints due to MicroBooNE's smaller detector mass.

Constraints on operators involving a tau neutrino would require atmospheric neutrinos or oscillated long-baseline $\nu_\tau$ measurements and are therefore beyond the scope of the current work.

\subsection{Light mediators and MiniBooNE excess}
\label{sec:MB}
As we saw in the previous sections, and shown in Fig.~\ref{fig:atomic_diffxs} and Fig.~\ref{fig:atomic_diffangle}, coherent $\nu$IIP scattering in the heavy dimension-7 EFT generically produces very forward-going final state photons. In contrast, the MiniBooNE low energy excess~\cite{MiniBooNE:2018esg,MiniBooNE:2020pnu} is characterized by a somewhat broader distribution in the cosine of the outgoing photon angle with respect to the beam axis, $\cos\theta_\gamma$, rather than being sharply peaked at  $\cos\theta_\gamma \simeq 1$. Accommodating these excess kinematics therefore requires a different momentum-transfer structure in the scattering amplitude than that of the pure contact interaction. 

A natural way to achieve this is to realize the polarizability operator through the exchange of a light mediator whose mass lies below that of the outgoing sterile state. In this case the amplitude acquires a nontrivial propagator factor, schematically $(q^2 - m_\phi^2)^{-1}$, which, in conjunction with the massive sterile state, softens the extreme forward enhancement of the contact limit and redistributes events to larger angles. As we show below, this light-mediator realization of the active–sterile polarizability can reproduce both the energy and angular distributions of the MiniBooNE excess while remaining consistent with other experimental constraints.

With this motivation in mind, we study a realization of this operator with a light (pseudo-)scalar $\phi$ coupled to neutrinos and to the photon field strength tensor,
\begin{align}
\mathscr{L}\supset 
\frac{g_{\phi\gamma}}{4}\phi F^{\mu\nu}\widetilde{F}_{\mu\nu}
+\frac{1}{2}c_\nu^{ij}(\overline{N}_{i}\nu_{j}) \phi+\text{h.c.} ,
\label{eq:fulllag}
\end{align}
where $g_{\phi\gamma}$ controls the axion-like coupling to photons and $c_\nu^{ij}$ parametrizes the couplings between active and sterile neutrinos. Such couplings, to the best of our knowledge, have not been studied in the literature, although related scenarios have been considered. For instance, the analysis with the coupling between axion-like particles and sterile neutrinos was conducted in \cite{Alves:2019xpc,deGiorgi:2022oks,Abdullahi:2023gdj}, the coupling between axion-like particles and active neutrinos was analyzed in \cite{Bonilla:2023dtf} and the couplings between Majorons and active neutrinos were discussed in \cite{Heeck:2019guh}.

\begin{figure*}[!t]
    \centering
    \includegraphics[width=0.45\linewidth]{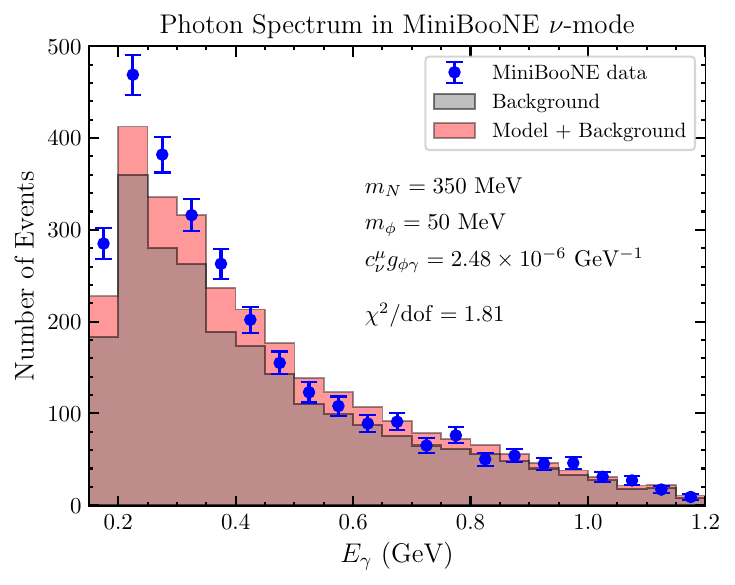}
     \includegraphics[width=0.45\linewidth]{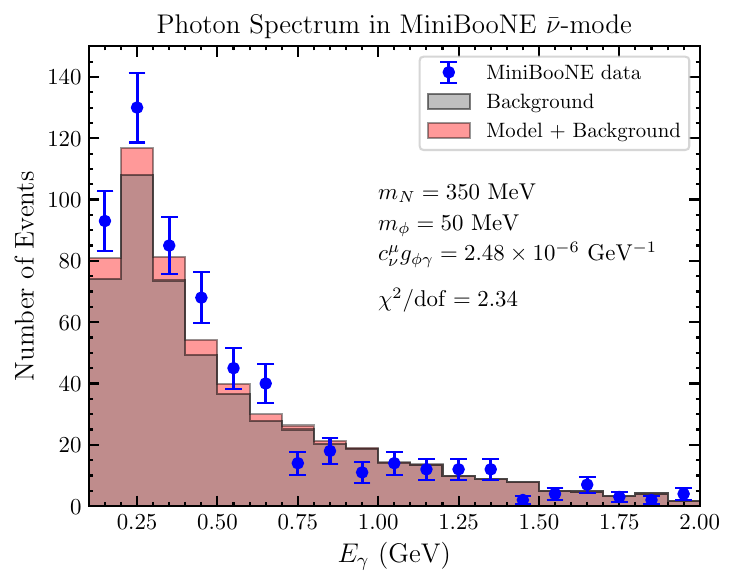}
     \includegraphics[width=0.45\linewidth]{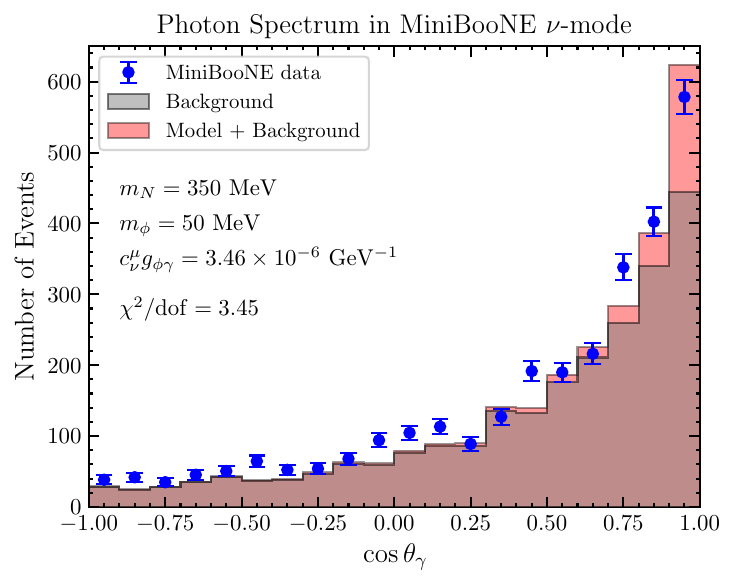}
     \includegraphics[width=0.45\linewidth]{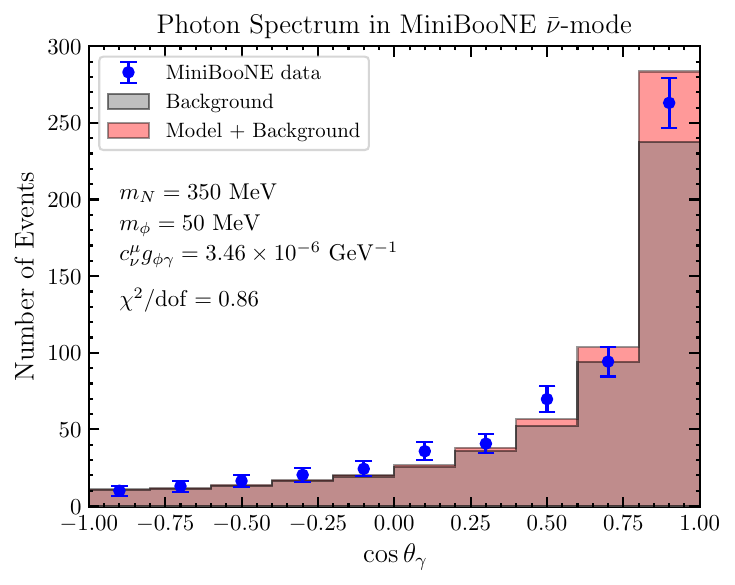}
    \caption{Photon energy (top) and angular distribution (bottom) spectrum from the MiniBooNE neutrino (left) and antineutrino (right) run \cite{MiniBooNE:2018esg,MiniBooNE:2020pnu}. The histograms shows the best-fit model prediction for the benchmark point $m_\phi = 50$ MeV and $m_N=350 $ MeV.  }
    \label{fig:MB_histos}
\end{figure*}

\begin{figure}
    \centering
    \includegraphics[width=0.45\textwidth]{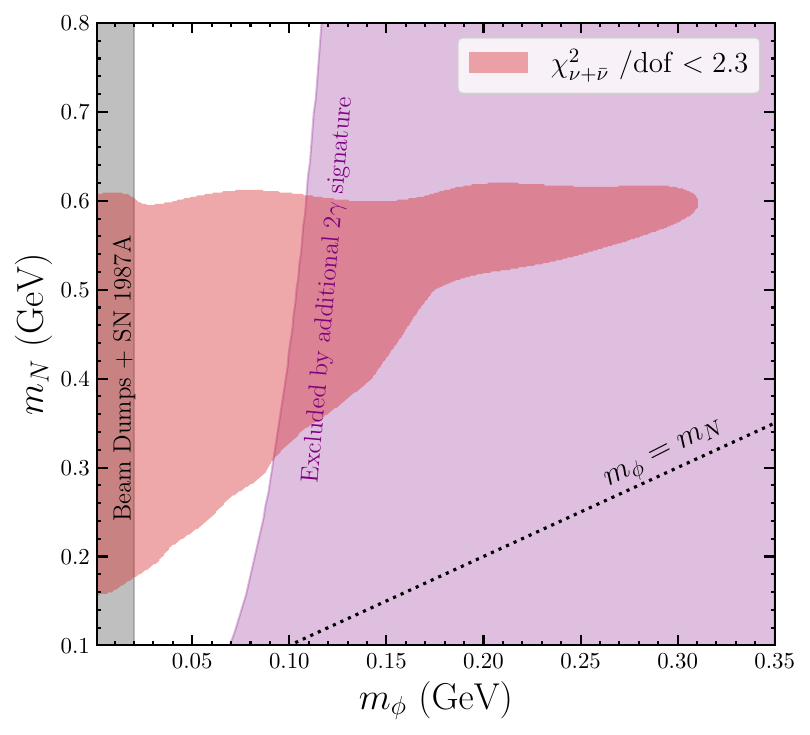}
    \caption{Allowed parameter space in the plane of the mediator mass $m_\phi$ and sterile-neutrino mass $m_N$ that provides an acceptable fit ($\chi^2/{\rm dof} < 2.3$) to the MiniBooNE anomaly (red). The region at very small $m_\phi$ is excluded by cosmological, astrophysical, and beam-dump constraints~\cite{Sabti:2021reh,Lucente:2020whw,Fiorillo:2022cdq,Bansal:2022zpi,Dolan:2017osp}. The purple shaded region represents the part of parameter space in which the outgoing $N$ typically decays to $N \to \nu \phi \to \nu \gamma\gamma$ inside the detector, so the resulting events fail the NC$1\gamma$ selection criteria and are not counted as single photon candidates.}
    \label{fig:MB_sol}
\end{figure}

The main contribution to the single photon final state is through the neutrino-induced inverse-Primakoff scattering ($\nu+{\cal A}\to N+\gamma+{\cal A}$). However, the Lagrangian in Eq.~\eqref{eq:fulllag} also induces a neutral-current process in which the incoming neutrino scatters off the target via $Z$ exchange and radiates a
$\phi$ from the external neutrino leg, followed by $\phi \to \gamma\gamma$, leading to $\nu + {\cal A} \to N + 2\gamma + {\cal A}$. This $Z$-mediated channel is however suppressed relative to $\nu$IIP by the extra weak
coupling and a heavy propagator, as well as by the three-body phase space. Moreover, the $2\gamma$ final state mimics a single photon like ring only when $\phi$ is highly boosted so that the two photons overlap within the Cherenkov angular resolution, which occurs only in a small fraction of phase space. As a result, after accounting for the reduced selection efficiency, the contribution of the $2\gamma$ channel to the NC$1\gamma$-like signal is negligible, and in what follows we focus on the $\nu$IIP channel as the dominant contribution.

Second, the non-observation of coherent mono-photon production at MicroBooNE~\cite{MicroBooNE:2025rsd} can, in principle, be utilized to set a weak constraint on coherent scattering that mimics the relevant amplitudes in the SM ~Ref.~\cite{PhysRevC.89.015503}. However, coherent scattering in the $\nu$IIP process via the light mediator model of Eq.~\ref{eq:fulllag} has a rather different amplitude structure, setting apart the two-dimensional $(E_\gamma, \cos\theta_\gamma)$ correlation from the SM coherent template used in Ref.~\cite{MicroBooNE:2025rsd}, corresponding to a greater inefficiency in the event selection. The $\nu$IIP coherent signal in presence of the light-mediator can therefore satisfy this bound, although deriving a precise constraint would require a dedicated reanalysis with an appropriate signal template and should motivate further study with future SBN data. In addition, incoherent scattering is subdominant in the region of parameter space we consider, ensuring that the final state contains sufficiently soft hadronic activity in order to be equivalent to a $1\gamma0p$ (one photon, zero proton) topology, consistent with the $1\gamma1p$ channels so far~\cite{MicroBooNE:2025ovj,MicroBooNE:2025ntu}.

From a fit to the photon energy and angular spectrum in the neutrino and antineutrino run of MiniBooNE, we find a best fit point in our model (see Fig.~\ref{fig:MB_histos}) with $m_N=350$ MeV, $m_\phi=50 $ MeV, and the product of couplings $c_\nu^\mu g_{\phi\gamma}=2.48\times 10^{-6}~\text{GeV}^{-1}$.  We find that this benchmark provides a better description of the data than the background-only hypothesis reported in Ref.~\cite{MiniBooNE:2020pnu}. The $\chi^2/\text{dof}$ for the photon energy and angular spectra in both runs is around 2, which is around 4 with only background.
Indeed,  in the neutrino mode our model substantially increases the number of events in the first four photon energy bins between $E_\gamma\in [0.15,0.4]$ GeV bringing it in better agreement with the observed data. In the antineutrino mode, our model leads to a smaller increase in the number of events in the first four excess bins. Nevertheless, the predicted number of signal and backgrounds events now agrees at the 1-1.5$\sigma$ level with the observed number of events. At photon energies above $E_\gamma\gtrsim 0.4$ GeV the data in the neutrino mode agrees mostly with the background only hypothesis, in the antineutrino mode the data presents deficit fluctuations in several bins. At these larger photon energies our model predicts only a small number of events in each bin, keeping the good agreement between data and background. For the photon angular spectrum, the model predicts only additional events in the forward direction with $\cos\theta_\gamma>0.75$ in both the neutrino and antineutrino mode, leading to an improved agreement between data and background plus signal hypothesis. In the very forward bin, our model actually leads to number of events compatible with the upper $1\sigma$ uncertainty on the data. 
In the bins with $\cos\theta_\gamma<0.75$, the data agrees with the background only hypothesis,  our model predicts only very few additional events, thereby not worsening the agreement.  

We note that the angular spectrum analysis prefers a slightly larger value of the coupling constant than the energy spectrum analysis. This can be explained as an increased coupling strength leads to too many events in the higher energy bins where the data fits the background well, therefore the energy spectrum analysis prefers a slightly smaller value of the coupling strength. The angular spectrum has some overfluctuations in the in some forward bins in the neutrino mode, so a slightly larger coupling is preferred.

Similar to other proposed explanations, the proposed model fits the antineutrino data slightly better than the neutrino data \cite{Fischer:2019fbw,Dutta:2021cip}. The best-fit point in our model can be tested at SBND and ICARUS with around $\mathcal{O}(10^{20})$ POT, depending on the experimental details, and are already situated to do so~\cite{SBND:2025lha,ICARUS:2024oqb}. 

Our mechanism differs from the eV-scale sterile neutrino scenario put forward to explain the MicroBooNE results. The eV sterile neutrino scenario  relies on active-sterile mixing with referred values which are in tension with results from atmospheric neutrino experiments, long-baseline experiments, and reactor neutrino experiments as well as cosmological observables \cite{Dentler:2018sju,Boser:2019rta}. Our scenario avoids these constraints altogether as we don't rely on the production of the sterile neutrinos via oscillations but instead they get produced in the scattering of the active neutrino in the detector together with a photon. These differences in the underlying mechanism to explain the anomaly leads to a heavier preferred sterile neutrino masses and hence allows for tests of this scenario not only at oscillation experiments (as it is the case for the sterile neutrino solution) but more generally in neutrino scattering experiments.

In Fig.~\ref{fig:MB_sol} we show the region in the $(m_\phi, m_N)$ plane that provides a good fit to the MiniBooNE data, defined by $\chi^2_{\nu+\bar{\nu}}/\text{dof}<2.3$. We find viable solutions for $180~\text{MeV} \lesssim m_N \lesssim 600~\text{MeV}$ and $m_\phi \lesssim 350~\text{MeV}$, with typical best-fit values of the coupling product around $c_\nu^\mu g_{\phi\gamma} \approx 10^{-6}$ GeV$^{-1}$. The gray vertical band indicates parameter space excluded by a combination of supernova~SN~1987A~\cite{Lucente:2020whw,Fiorillo:2022cdq,Bansal:2022zpi} and beam-dump constraints~\cite{Dolan:2017osp}, which requires $g_{\phi \gamma}$ to be extremely small, of order $10^{-11}\ {\rm GeV}^{-1}$ for $m_\phi < 20$ MeV. For such tiny  $g_{\phi\gamma}$, reproducing the fitted product $c_\nu^\mu g_{\phi\gamma}$ would require $c_\nu^\mu$ to be non-perturbatively large, and we therefore discard this region. In addition, CMB and BBN observations constrain the mass of a light neutrino-philic mediator to be above a few MeV~\cite{Sabti:2021reh}, a requirement that is comfortably satisfied throughout the parameter space displayed in Fig.~\ref{fig:MB_sol}.

Let us now turn to the possibility of $N$ decaying inside the detector after  its production.  Unlike the case of the active-sterile polarizability arising from an EFT with heavy degrees of freedom as discussed in \S~\ref{sec:pheno}, where $N \to \nu \gamma \gamma$ proceeds through a contact interaction, here the  mediator $\phi$ is lighter than $N$ allowing  for the two body prompt decay $N \to \nu \phi$ followed by $\phi \to \gamma \gamma$. 
This leads to a ``double-bang''-like signature, an initial $\nu$IIP interaction that produces $N$ and a photon, and a secondary vertex where $N$ decays to $\nu$ and a di-photon state. This signature should be highly unique and depends on the kinematics, the detector volume, and the model parameter space.  
At MiniBooNE, such a decay mode would ordinarily be problematic, leading to a secondary pair of Cherenkov rings from $\phi \to \gamma \gamma$ in addition to the primary mono-photon signal, provided that the $\phi$ decay length is shorter than the detector size. If $N$ decays promptly and with modest energy, the resulting $\phi$ can be too soft for its two photons to exceed the visible-energy threshold ($E_{\rm vis} \simeq 140$~MeV), so that the secondary di-photon state is effectively invisible. Conversely, for more energetic $N$ in the high-energy tail of the spectrum would contribute an addition di-photon pair in the event, thus failing the single shower selection criteria and thereby reducing the overall efficiency of the signal.

However, these aforementioned possibilities are coupling- and mass-dependent; therefore, only a subset of the parameter space in Fig.~\ref{fig:MB_sol} (which profiles over the couplings, shown by the purple shaded region) will partially satisfy the requirement that $N \to \nu \phi(\to \gamma\gamma)$ is either sufficiently boosted or soft enough to avoid a clear $1\gamma$ + $2\gamma$ signatures. We find that this requirement is generically easier to satisfy for $m_\phi \lesssim 100$ MeV, where the $1\gamma$ + $2\gamma$ signature is inefficient for the characteristic boost factors in the experiment. Relaxing this requirement, this potential double-bang feature presents a unique and complementary probe at other intensity-frontier experiments with large detector volumes and different beam energies, as well as in astrophysical environments or early universe cosmological observables~\cite{Bansal:2022zpi} possibly connected to the discussion of sterile neutrino dark matter~\cite{DeGouvea:2019wpf}.

Finally, MicroBooNE has recently observed a mild excess of photon-like events \cite{MicroBooNE:2025ntu}. For the best-fit point which can explain the MiniBooNE data in our model, we expect around 70 events at MicroBooNE compared to $93\pm 22\pm35$ observed excess events.

\section{Model realizations of this operator}
\label{sec:models}
In this section, we briefly discuss two possible UV complete realizations of the active-sterile neutrino polarizability operator for the sake of completeness.

\subsection{Mixing with the sterile neutrino via a charged lepton loop}

Let us first illustrate how the polarizability operator can arise from the electroweak Lagrangian in the presence of sterile neutrinos that have mixings with the three active generations, and then evaluate whether a substantial phenomenological enhancement is possible. To generate the effective neutrino-photon coupling in this simple extension of the Standard Model, we rely on the neutrino couplings to charged fermions, and the charged fermions  coupling to photons. This generates the  effective neutrino-photon coupling at 1-loop. Below the electroweak scale, neutrino interactions with charged leptons are described by four-fermi theory
\begin{equation}
    \mathscr{L} \supset \frac{G_F}{2 \sqrt{2}} \overline{\nu}_\alpha \gamma^\mu (1 - \gamma^5) \nu_\alpha J_{\mu, weak}^\beta \, ,
\end{equation}
where $\alpha,\beta = e, \mu, \tau$ denote the lepton flavors and 
$J_{\mu,\text{weak}}^\beta$ is the corresponding weak charged-lepton current. In a simplified 3+1 framework with one heavy neutrino mass eigenstate $N\equiv \nu_4$, we express the flavor states $\nu_\alpha$ as $\nu_\alpha = U_{\alpha i}^* N$ where $U_{\alpha i}$ is the $4\times4$ neutrino mass mixing matrix. We can consider transitions from $\nu_\alpha$ to $N$ by expanding the 4-fermi Lagrangian to include the mass insertion,
\begin{equation}
    \mathscr{L}_{\alpha 4} = \frac{G_F}{2 \sqrt{2}} \overline{\nu}_e \gamma^\mu (1 - \gamma^5) U_{e 4}^* N J_{\mu, weak}^\beta \, .
\end{equation}

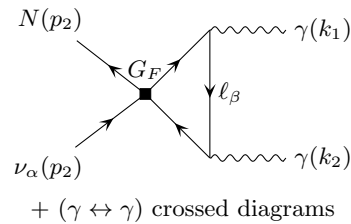
\begin{figure}[ht!]
    \centering
    \begin{tikzpicture}
              \begin{feynman}
         \vertex (c);
         \vertex [above right = 1.2cm of c] (l1);
         \vertex [below right = 1.2cm of c] (l2);
         \vertex [right = 1.0cm of l1] (f1) {\(\gamma(k_1)\)};
         \vertex [right = 1.0cm of l2] (f2) {\(\gamma(k_2)\)};
         \vertex [above left = 2.2cm of l2] (i1) {\(N (p_2)\)};
         \vertex [below left = 2.2cm of l1] (i2) {\(\nu_\alpha (p_2)\)};

         \node[draw, fill, rectangle, inner sep=2pt, label=\(G_F\)] at (c) {};
         
         \diagram* {
           (l1) -- [boson] (f1),
           (l2) -- [boson] (f2),
           (i2) -- [fermion1] (c) -- [fermion1] (l1) -- [fermion1, edge label=\(\ell_\beta\)] (l2) -- [fermion1] (c) -- [fermion1] (i1)
         };
        \end{feynman}
       \end{tikzpicture}

       + ($\gamma \leftrightarrow \gamma)$ crossed diagrams
    \caption{The leading 1-loop diagram contributing to the $2\nu2\gamma$ effective operator. A secondary diagram involving quark loops is sufficiently subleading to ignore.}
    \label{fig:loop-diag_4fermi}
\end{figure}

Below the weak scale, this interaction gives rise to a $\nu \nu \gamma \gamma$ vertex at one-loop through charged-lepton loops, as shown in Fig.~\ref{fig:loop-diag_4fermi}. A secondary diagram involving quark loops is subleading and will be ignored here. The computation, summarized in Appendix~\ref{app:ewpol}, leads to the following form of the polarizability operator
\begin{equation}
    \mathscr{L}_{\rm pol}^{\alpha 4} = \frac{G_F\alpha}{24 \pi \sqrt{2}}  m_4 U_{\alpha 4}
    \sum_\alpha\bigg(-\frac{I_\alpha}{m_\alpha^2} + \sum_{\beta \neq \alpha}  \frac{I_\beta}{m_\beta^2} \bigg) \overline{\nu}_4 \gamma^5 \nu_\alpha F \Tilde{F} \, ,
\end{equation}
where $m_\alpha$ is the charged lepton mass for $\alpha = e,\mu,\tau$ and $I_\alpha \equiv I(m_\alpha^2 / 2 k_1 \cdot k_2)$ is a loop integral defined in Eq.~\ref{eq:loop}. 

The dominant contribution in this realization always comes from the electron loop, which is the least suppressed by the factor $m_\alpha^{-2}$, i.e., $\propto I_e / m_e^2$. 
The loop integral itself quickly tends to zero in the large momentum limit beyond $2 k_1 \cdot k_2 \gtrsim m_\alpha^2$. In the case of 
$\nu_\alpha \mathcal{A}\to \mathcal{A}\,N \gamma$ scattering with $\alpha = e$, the $W^\pm$ and $Z^0$ diagrams both contribute at the four-Fermi level, while for $\alpha \neq e$, only the $Z^0$ diagram contributes.

By itself, this model realization is therefore difficult to probe at accelerator-based experiments: for neutrino energies well above the MeV scale, the loop factors $I_\beta$ are small for all charged leptons, and the polarizability is highly suppressed. Reactor antineutrinos and solar neutrinos are thus the most suitable probes, especially with large detectors capable of detecting MeV-energy photons from $\bar{\nu}_e \mathcal{A} \to \mathcal{A}~\bar{N} \gamma$ scattering. However, we find that the mixing angles  $U_{\alpha 4}$ much smaller than $10^{-1}$ are challenging to probe.  

It is important to note that the enhancements to this realization could arise in extension of the SM with new gauge interactions, for example with $W^\prime$ or $Z^\prime$ bosons that couple to charged leptons and neutrinos. However, as noted in Appendix~\ref{app:ewpol}, pure vector currents coupling to the charged lepton loop and two photons will vanish, requiring either chiral or pure axial vector couplings.

\subsection{Realization with a singly charged scalar}
To enable the generation of the neutrino polarizability operator at one loop,
we add a singly charged scalar singlet $S^+ \sim (1,1,1)$ and the chiral right-handed neutrinos $N_i$ to the SM. 
The relevant Yukawa interactions involving $S^+$ and $N_{i}$ are
\begin{align}
    \mathcal{L} \supset &
    + \frac{1}{2} f_{\alpha \beta}\, L_\alpha^{T} C i \sigma_2 L_\beta\, S^+
    + \lambda_{i\beta}\, \overline{N}_{i}^{\,c}\, e_{R\beta}\, S^+ + \text{h.c.} \, ,
\end{align}
The scalar potential is given by
\begin{equation}
    V \supset V_{\rm SM} 
    + m_S^2\, S^\dagger S 
    + \lambda_S\, (S^\dagger S)^2 
    + \lambda_{HS}\, (H^\dagger H)(S^\dagger S)\, .
\end{equation}
Here $f_{\alpha\beta}=-f_{\beta\alpha}$.
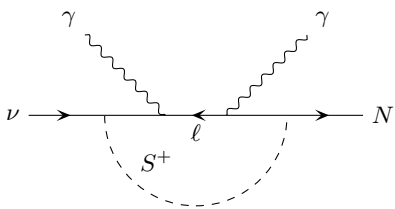
\begin{figure}[ht!]
    \centering
       \begin{tikzpicture}
              \begin{feynman}
         \vertex (c);
         \vertex [left = 1.2cm of c] (c1);
         \vertex [right = 1.2cm of c] (c2);
         \vertex [left = 0.4cm of c] (c1i);
         \vertex [right = 0.4cm of c] (c2i);
         \vertex [left = 1.0cm of c1] (fi) {\(\nu\)};
         \vertex [right = 1.0cm of c2] (ff) {\(N\)};
         \vertex [below = 1.2cm of c] (h);
         \vertex [above left = 1.5cm of c1i] (p1) {\(\gamma\)};
         \vertex [above right = 1.5cm of c2i] (p2) {\(\gamma\)};

         \diagram* {
           (fi) -- [fermion1] (c1),
           (c2) -- [fermion1] (ff),
           (c2) -- [fermion1, edge label=\(\ell\)] (c1),
           (c1) -- [scalar, out=270, in=180, edge label=\(S^+\)] (h)
                      -- [scalar, out=0, in=270] (c2),
            (c1i) -- [boson] (p1),
            (c2i) -- [boson] (p2)
         };
        \end{feynman}
        \end{tikzpicture}

    \caption{Typical one-loop diagram contributing to the neutrino polarizability from the scalar singlet $S^+$.}
    \label{fig:loop-charged}
\end{figure}
%%%
One can then draw the one-loop diagram that induces the effective operators 
$(\overline{\nu}_{L\alpha}N_{ i})\,F_{\mu\nu}\tilde F^{\mu\nu}$ and $(\overline{\nu}_{L\alpha}N_{ i})\,F_{\mu\nu} F^{\mu\nu}$ shown schematically in Fig.~\ref{fig:loop-charged}. Additional diagrams exist where the photons are emitted from the charged scalar.   
After integrating out the heavy scalar, and in the limit of vanishing external momenta and $m_S \gg m_{\beta}$ ($\beta = e, \mu, \tau$), the Wilson coefficients for the CP-even where CP stands for charge conjugation and parity, read as
\begin{align}
    \frac{C_{\alpha i}}{\Lambda^3} 
    &\simeq 
    \frac{e^2}{16\pi^2} \frac{1}{6}
    \sum_{\beta} (f_{\alpha\beta}\lambda_{i\beta})\,
    \frac{m_{\beta}}{m_S^4}\, ,
    \label{eq:C7-even-multi}
\end{align}
The CP-odd term ($F_{\mu\nu}\tilde F^{\mu\nu}$) originates from the imaginary part of the loop amplitude and is therefore proportional to  $\mathrm{Im}(f_{\alpha\beta}\lambda_{i\beta})$ and contains extra factor $(1/2\ \ln (m_S^2/m_\beta^2 -1)$ instead of factor $1/6$.  The same loop topology that generates the Rayleigh operator also induces  the dipole operator $\mu_{\alpha i}$ (upon removing one external photon)  and the radiative Dirac mass $(m_D)_{\alpha i}$ (upon removing both photons).   In the limit $m_S \gg m_{\beta}$, the corresponding contributions are
\begin{align}
    \mu_{\alpha i} &\simeq 
    -\,\frac{e}{16\pi^2}\,\frac{1}{6}
    \sum_\beta f_{\alpha\beta}\lambda_{i\beta}\,
    \frac{m_{\beta}}{m_S^2}\, , 
    \label{eq:dipole-multi}\\[6pt]
    (m_D)_{\alpha i} &\simeq
    \frac{1}{16\pi^2}\sum_\beta
    f_{\alpha\beta}\lambda_{i\beta}\,m_{\beta}\,
   \frac{m_{\beta}^2}{m_S^2}
    \ln\!\frac{m_{\beta}^2}{m_S^2}\, .
    \label{eq:diracmass-multi}
\end{align}
Because $f_{\alpha\beta}=-f_{\beta\alpha}$, the terms with $\beta=\alpha$ vanish, implying that at least two distinct lepton flavors are required for a nonzero contribution.  This mechanism thus provides a simple  framework in which neutrino Dirac masses, dipole couplings, and polarizability arise radiatively through the same set of interactions.

\subsection{Other options}
Finally,  other UV realizations of the neutrino polarizability operator are also possible. For example, the pseudo-scalar mediator could be identified as a dark pion that mixes with the SM pion \cite{Berryman:2017twh}, which in turn decays into two photons \cite{Bell:1969ts,Adler:1969gk}. We leave a detailed exploration of such scenarios and their signatures for future work.

\section{Conclusions}
\label{sec:conclusions}
Testing new interactions of neutrinos beyond the Standard Model could provide insights into new physics extensions. Among the most popular new physics extensions is the addition of a sterile neutrino, which could also participate in new neutrino interactions. In this manuscript, we have focused on a particular new neutrino interaction with two photons and a sterile neutrino, namely the active-sterile neutrino polarizability operator. The signature of this new operator is two-fold: it leads to a new contribution to neutrino-nucleus neutral current scattering with a mono-photon in the final state, and it leads to a new decay of the sterile neutrino which does not involve its mixing with the active neutrinos.
Compared to the neutrino polarizability operator involving only active neutrinos, the energy spectrum of the outgoing photon  in a neutral current scattering process gets softened and shifted toward lower energies depending on the sterile neutrino mass. 

We have derived constraints on the effective operator from existing experimental data at NOMAD and MiniBooNE, and illustrated the potential reach at future experiments. A UV realization of this operator with a light pseudo-scalar is of particular interest and can explain the MiniBooNE low-energy excess for mediator masses of around 50 MeV and sterile neutrino masses of around 350 MeV with coupling products on the order of $c_\nu^\mu g_{\phi\gamma}\approx 3\times 10^{-6}~\text{GeV}^{-1}$. Upcoming results from the SBN program will provide more information about the viability of a mono-photon explanation of the excess, and its fine-grained kinematic nature, while spallation sources or reactors may utilize alternative neutrino flux flavor compositions to test the consistency of a polarizability hypothesis in a complementary way.

\section*{Acknowledgments}
We thank Sam Carey, Andr\'{e} de G\^{o}uvea, Matheus Hostert, and Zahra Tabrizi for the helpful theoretical discussions.
We would like to thank the Center for Theoretical Underground Physics and Related Areas (CETUP*) and the Institute for Underground Science at Sanford Underground Research Facility (SURF) for providing a conducive environment during the 2025 summer workshop.
JG acknowledges support by the U.~S.~ Department of Energy Office of Science under award number DE-SC0025448. The work of AT is supported in part by the U.S. DOE grant DE-SC0010143.

\appendix

\section{The Polarizability operator in the \texorpdfstring{$\nu$SM}{NSM} }
\label{app:ewpol}

The following matching of the amplitude is exactly like the approach taken by Crewther, Finjord, and Minkowski~\cite{Crewther:1981wh}, which we reiterate here for clarity. We express the leading contribution to the $\nu\nu\gamma\gamma$ amplitude with momenta $\nu_\alpha(p_1), N(p_2)$ and outgoing photon momenta $k_1, k_2$ and spins $s_1, s_2$ in terms of $\bra{\gamma(k_1,s_1),\gamma(k_2,s_2)}$ creation amplitudes sourced by the axial-vector current;
\begin{align}
    \mathcal{M}_{\nu_\alpha N \gamma \gamma} &= i\frac{G_F U_{\alpha 4}}{2\sqrt{2}} \bar{u}_4(p_2) \gamma^\sigma (1 - \gamma^5) u_\alpha(p_1) \nonumber \\
     \times &\bigg( -\big\langle{\gamma(k_1,\sigma_1),\gamma(k_2,\sigma_2) \big|J_{\sigma,5}^{\alpha}  \big| 0}\big\rangle \nonumber \\
    & + \sum_{\beta \neq \alpha} \big\langle{\gamma(k_1,\sigma_1),\gamma(k_2,\sigma_2) \big| J_{\sigma,5}^{\beta} \big| 0}\big\rangle \bigg)  \, ,
\end{align}
where the two terms inside the large parentheses account for the summation of both neutral and charged current contributions to the charged fermion loop. Since there is no vector current anomaly, only the axial-vector current $J_{\mu,5}^\alpha \equiv \bar{\ell}_\alpha \gamma_\sigma \gamma^5 \ell_\alpha$ contributes.

We use the result
\begin{align}
&\big\langle{\gamma(k_1,\sigma_1),\gamma(k_2,\sigma_2) \big|J_{\sigma,5}^{\alpha}  \big| 0}\big\rangle \nonumber \\
&= -i\frac{\alpha}{6\pi} (k_1 + k_2)_\sigma \epsilon_{\mu\nu\lambda\delta} \varepsilon^\mu_1 \varepsilon^\nu_2 k_1^\lambda k_2^\delta \frac{I(\tau_\alpha)}{m_\alpha^2} \, ,
\end{align}
where $\varepsilon_{1}^{\mu}, \varepsilon_{2}^{\nu}$ are the $\gamma$ polarization vectors and $\tau_\alpha = m_\alpha^2 / (2 k_1 \cdot k_2)$ is the argument in the loop integral,
\begin{equation}
\label{eq:loop}
    I(\tau_\alpha) \equiv 24 \tau_\alpha \int_0^1 dx \int_0^{1-x} \frac{x y}{xy - \tau_\alpha} dy \, .
\end{equation}
At energy scales below $m_\alpha^2$, e.g. as $\tau_\alpha \to \infty$, $I_\alpha \to -1$, and otherwise the loop factor vanishes in the high energy limit.

Putting the pieces together, the matrix element reads
\begin{align}
    &\mathcal{M}_{\nu_\alpha N \gamma \gamma} = -i\frac{\alpha G_F U_{\alpha 4}}{12\pi\sqrt{2}} \bar{u}_4(p_2) \gamma^\sigma \gamma^5 u_\alpha(p_1)  \nonumber \\
    &\times  (k_1 + k_2)_\sigma \epsilon_{\mu\nu\lambda\delta} \varepsilon^\mu_1 \varepsilon^\nu_2 k_1^\lambda k_2^\delta \bigg( - \frac{I(\tau_\alpha)}{m_\alpha^2} + \sum_{\beta \neq \alpha} \frac{I(\tau_\beta)}{m_\beta^2}\bigg) \, . 
\end{align}
Due to momentum conservation, $(k_1 + k_2)_\sigma = (p_1 + p_2)_\sigma$, which can then be absorbed next to the $\bar{u}_4$ and $u_\alpha(p_1)$ spinors via the Dirac equation to pull out factors of the neutrino masses, for which we neglect contributions from the light neutrino masses for the moment. Matching the factor of $\epsilon_{\mu\nu\lambda\delta} \varepsilon^\mu_1 \varepsilon^\nu_2 k_1^\lambda k_2^\delta$ to the CP-odd operator $F\Tilde{F}/4 = \epsilon_{\mu\nu\lambda\delta} F^{\mu\nu}F^{\lambda\delta}/4$, for the electromagnetic field-strength tensor $F$, leads us to an effective Polarizability operator
\begin{equation}
    \mathscr{L}_{\rm pol}^{\alpha 4} = \frac{G_F\alpha}{24 \pi \sqrt{2}}  m_4 U_{\alpha 4}
    \sum_\alpha\bigg(-\frac{I_\alpha}{m_\alpha^2} + \sum_{\beta \neq \alpha}  \frac{I_\beta}{m_\beta^2} \bigg) \overline{\nu}_4 \gamma^5 \nu_\alpha F \Tilde{F} \, .
\end{equation}
In the case of active neutrinos at energy scales below $m_e^2$, we point out that the dominant contribution to the dimension-7 coefficient is proportional to the heaviest neutrino mass,
\begin{align}
    \frac{C_{ij,\rm max}}{\Lambda^3} &\to \frac{G_F \alpha}{24 \pi \sqrt{2} m_e^2} m_{\nu} \nonumber \\
    &\sim 3 \times 10^{-12} \bigg(\frac{m_\nu}{0.1\text{ eV}}\bigg) \, [\text{GeV}^{-3}] \, ,
\end{align}
or around 3 orders of magnitude below the foreseeable reach we find in Fig.~\ref{fig:constraint_wilson}.

\section{Details of experiments}

\subsection{JUNO--TAO}

JUNO--TAO is a ton-scale Gd-doped liquid scintillator (GdLS) detector located at a short baseline ($L \sim 30$–$40$ m) from one of the $P_{\rm th}\simeq 4.6~{\rm GW_{th}}$ Taishan reactor cores and serves as a high-resolution near detector for JUNO~\cite{JUNO:2020ijm}. The central spherical acrylic vessel contains 2.8~tons of GdLS, with a 1~ton fiducial target corresponding to 
$N_T \simeq 1.2\times10^{28}$
free carbon atoms, which are approximately the dominant coherent scattering target. We use the reactor $\bar\nu_e$ flux $\phi_{\rm tot}(E_\nu)$ at a baseline of $L=30$~m and a nominal IBD event rate of $\sim 2000$/day from Ref.~\cite{JUNO:2020ijm}, restricting to $E_\nu \in [1.8,8]$~MeV. No dedicated single-photon search exists, so we treat JUNO--TAO in projected-sensitivity mode, assuming that photons are reconstructed as electromagnetic showers with the same response as positrons, a constant photon efficiency $\epsilon_\gamma \simeq 50\%$ for $E_\gamma > 0.9$~MeV.

\subsection{PROSPECT-II at HFIR}

PROSPECT-II is a short-baseline detector at the 85~MW$_\text{th}$ HFIR research reactor, with baselines $L \simeq 7$–12~m from the compact, highly-enriched $^{235}$U core~\cite{PROSPECT:2021jey}. In phase II, a future run at a GW-scale low enriched uranium (LEU) reactor could also be possible. The detector consists of a segmented $^6$Li-doped liquid-scintillator (LiLS) target with 4.8~tons of active mass, and we approximate the coherent scattering rate with
$N_T \simeq 6\times10^{28}$ free carbon atoms.
We take the $\bar\nu_e$ flux at $L=7.9$~m from Ref.~\cite{PhysRevC.83.054615,TEXONO:2009knm} and nominal 2-year exposure (336 reactor-on days), which yields $\sim 4 \times 10^5$ IBD candidates with an overall IBD efficiency of about 40\%. For our polarizability signal we assume single photons are reconstructed in the same prompt window as IBD events, with the same assumed constant efficiency for photon detection, and neglect any angular dependence. We then include PROSPECT-II as a binned $E_\gamma$ measurement in the combined $\chi^2$.

\subsection{MiniBooNE}
   
MiniBooNE was a short baseline experiment that used a pure mineral-oil (CH$_2$) Cherenkov detector, which simultaneously served as neutrino target and scintillating medium. Because the detector cannot distinguish an electron from a single photon shower, electron–neutrino interactions and NC$1\gamma$ events produce indistinguishable signatures. The beam is dominated by $\nu_\mu$ and $\bar{\nu}_\mu$ and peaks at neutrino energies of $E_\nu \sim 500$~MeV, corresponding to a typical momentum transfer of $q^2 \sim -2~\text{GeV}^2$. The total neutrino flux $\phi_{\rm tot}(E_\nu)$ in units of [$\nu$/POT/GeV/$\text{cm}^2$] for both neutrino and antineutrino running is taken from Ref.~\cite{MiniBooNE:2008hfu}. The data sets used in our analysis correspond to exposures of $18.75 \times 10^{20}$ POT (neutrino mode) and $11.27 \times 10^{20}$ POT (antineutrino mode). For the signal prediction we adopt the photon detection efficiency $\epsilon(E_\gamma)$ from Ref.~\cite{MiniBooNE:2012maf}, impose a minimum reconstructed photon energy of $E_\gamma > 100~\text{MeV}$~\cite{MiniBooNE:2020pnu,MiniBooNE:2018esg}, and take the number of target carbon nuclei in the fiducial volume to be $N_T = 4.6 \times 10^{31}$.

\subsection{NOMAD}

The NOMAD experiment carried out a dedicated search for single-photon events using a drift-chamber target with a fiducial mass of 2.7~tons. The target material was a mixture of carbon (64\%), oxygen (22\%), nitrogen (6\%), and hydrogen (5\%), corresponding to an effective atomic mass $A \simeq 12.8$~\cite{NOMAD:2011gyy}. The neutrino beam spectrum used in our simulation is taken from Ref.~\cite{NOMAD:2003owt} and spans $E_\nu \simeq 5$--$100$~GeV, for a total exposure of $5.1 \times 10^{19}$~POT.

In modeling NOMAD’s sensitivity to our signal, we adopt a constant single-photon detection efficiency of 8\%~\cite{NOMAD:2011gyy}, applied uniformly across both photon energy $E_\gamma$ and angle $\cos\theta$. We also implement the photon energy–angle selection
\begin{equation}
    E_\gamma \bigl(1 - \cos\theta\bigr) \le 0.05\,,
\end{equation}
where $\theta$ is the angle of the photon with respect to the beam direction. Finally, we assume that all signal events satisfy the PAN requirement defined in Ref.~\cite{NOMAD:2011gyy}, which is based on the fraction of neutral energy deposited in the electromagnetic calorimeter, and therefore treat its impact as an overall acceptance factor already included in the quoted efficiency.

\subsection{SBND and ICARUS}

SBND is located 110~m downstream of the BNB target, while ICARUS sits at a baseline of 600~m; both sample the same Booster Neutrino Beam (BNB) as MicroBooNE. For SBND we use the BNB flux prediction from Ref.~\cite{sbndflux}, and obtain the BNB flux at ICARUS by rescaling the SBND flux by the geometric factor $(110/600)^2$. Neither SBND nor ICARUS has yet reported a dedicated single-photon search. Accordingly, we treat both experiments in a projected-sensitivity mode, assuming a constant photon-detection efficiency of 10\%, consistent with the efficiency used for MicroBooNE, and impose a minimum photon-energy requirement of $E_\gamma > 20~\text{MeV}$ to suppress low-energy backgrounds.

For the detector masses, we take a fiducial volume of 112~tons of liquid argon for SBND, corresponding to $N_T = 9.4 \times 10^{28}$ argon nuclei, and 476~tons for ICARUS, corresponding to $N_T = 4.0 \times 10^{29}$. For ICARUS we assume an exposure of $3.8 \times 10^{20}$~POT on the BNB, based on current running~\cite{DiNoto:2024nwm}. Note that 
ICARUS also obtains a neutrino flux from the NuMI beam which has delivered $2.28 \times 10^{20}$ POT for antineutrino mode in RHC \cite{DiNoto:2024nwm}. We expect the constraints from the NuMI flux to be comparable yet slightly weaker than the constraints obtained from the BNB.

For SBND, which has only recently begun taking data and does not yet have a mono-photon analysis, we adopt a benchmark exposure of $6.6 \times 10^{20}$~POT and apply the same uniform 10\% efficiency across all kinematic bins. Modest variations in the assumed POT exposures do not qualitatively impact our results.

\subsection{DUNE}

The DUNE near detector complex will feature excellent energy reconstruction and particle-identification capabilities, including the ability to distinguish photons from electrons. For our purposes we focus on the liquid-argon near detector (LAr ND), located 574~m from the LBNF target. We assume a fiducial mass of 67~tons, which corresponds to approximately $N_T = 5.6 \times 10^{28}$ argon nuclei in the active volume. We use the DUNE flux prediction and photon reconstruction efficiency from the DUNE TDR~\cite{DUNE:2020ypp}. As in the SBN analysis, we impose a minimum photon energy cut $E_\gamma > 20~\text{MeV}$ and consider an exposure of $1.1 \times 10^{21}$~POT per year. In evaluating the signal, we restrict the incoming neutrino energy to the range $E_\nu \in [0.5, 10]~\text{GeV}$, which covers the bulk of the DUNE LBNF flux relevant for our study.

\bibliography{main}

\end{document}